\documentclass[preprint2,dvipsnames,x11names]{aastex631}
\usepackage{amsmath}
\usepackage{physics}
\usepackage{hyperref}
\hypersetup{colorlinks=true, citecolor=blue, urlcolor=blue, linkcolor=blue}
\usepackage{xcolor}
\usepackage{enumerate}
\setcitestyle{notesep={ }}

\definecolor{ao(english)}{rgb}{0.0, 0.5, 0.0}

\def\atag{\refstepcounter{equation}\tag{\arabic{equation}}}
\begin{document}
\title{A Model for Pair Production Limit Cycles in Pulsar Magnetospheres}

\author[0009-0003-9478-380X]{Takuya Okawa}
\affil{Physics Department and McDonnell Center for the Space Sciences, Washington University in St. Louis; MO, 63130, USA}
\author[0000-0002-4738-1168]{Alexander Y. Chen}
\affil{Physics Department and McDonnell Center for the Space Sciences, Washington University in St. Louis; MO, 63130, USA}

\correspondingauthor{Takuya Okawa}
\email{o.takuya@wustl.edu}

\begin{abstract}
    It was recently proposed that the electric field oscillation as a result
      of self-consistent $e^{\pm}$ pair production may be the source of coherent
      radio emission from pulsars. Direct Particle-in-Cell (PIC) simulations of
    this process have shown that the screening of the parallel electric
      field by this pair cascade manifests as a limit cycle, as the
      parallel electric field is recurrently induced when pairs produced in the
      cascade escape from the gap region. In this work, we develop a simplified
    time-dependent kinetic model of $e^{\pm}$ pair cascades in pulsar
    magnetospheres that can reproduce the limit-cycle behavior of pair
      production and electric field screening. This model includes the
      effects of a magnetospheric current, the escape of $e^{\pm}$, as well as
      the dynamic dependence of pair production rate on the plasma density and
      energy. Using this simple theoretical model, we show that the power
      spectrum of electric field oscillations averaged over many limit cycles is compatible with the observed
      pulsar radio spectrum.
\end{abstract}


\section{Introduction}
\indent
Pulsars are rapidly rotating, highly magnetized neutron stars that produce coherent radio emission with enormous brightness temperature~\citep[see e.g.][for a review]{doi:10.1146/annurev-astro-052920-112338}. Very quickly after its discovery, it was realized that the magnetic field near the pulsar surface can be strong enough to ignite a QED $e^\pm$ pair cascade~\citep{1971ApJ...164..529S}. It was believed that pulsars can fill their surroundings with plasma through this $e^\pm$ pair production process, screening the electric field $E_{\parallel}$ along the magnetic field, creating a smooth force-free magnetosphere~\citep{1999ApJ...511..351C,2006ApJ...648L..51S}. The regions in the magnetosphere with unscreened $\mathbf{E}\cdot\mathbf{B}$ are called ``gaps'' and they are the main locations for pair production activity~\citep[see e.g.][]{1975ApJ...196...51R,1983ApJ...266..215A,1986ApJ...300..500C}.

Arguably, the most important pair-producing gap is located at the pulsar polar
cap, since it supplies the plasma on open field lines that is believed to be the
source of coherent radio emission~\citep{1971ApJ...164..529S}. \citet{2008ApJ...683L..41B} demonstrated
theoretically that the pair production process at the polar cap must be
inherently time-dependent when the magnetospheric current is spacelike, and \citet{Levinson:2005fa} showed that this process
tends to produce large-amplitude oscillations of the accelerating electric field. Numerical simulations performed by \citet{2010MNRAS.408.2092T}
demonstrated such oscillations from first-principles, and showed that pair production happens in quasiperiodic bursts. Subsequent 1D and 2D
Particle-in-Cell (PIC) simulations all showed the limit-cycle behavior of the
pair production process~\citep[e.g.][]{2013MNRAS.429...20T,2021ApJ...919L...4C},
and it was proposed that this pair-production oscillation may directly source
coherent radio waves~\citep{Philippov:2020jxu}. A better understanding of the
$e^\pm$ discharge physics may help us finally solve the decades-old puzzle of
the origin of pulsar radio emission.

Recently, it was also realized that the intense parallel electric field in the
gap region may be the strongest source of persistent oscillatory
$\mathbf{E}\cdot\mathbf{B}$ in the universe. Pseudoscalar particles such as
QCD axions interact with the electromagnetic sector through their coupling to
$\mathbf{E}\cdot\mathbf{B}$, therefore the spark gaps in the pulsar
magnetosphere may be one of the most promising regions of producing these
axion-like particles (ALPs)~\citep{Prabhu:2021zve}. The QCD axions and ALPs are physically motivated and form a
popular class of dark matter candidates~\citep{Preskill:1982cy,Abbott:1982af,Dine:1982ah}. Recent work by
\citet{Noordhuis:2022ljw} used the pair cascade process at pulsar polar caps to
derive novel constraints on axion signals. A better understanding of the complex
plasma physics behind $e^\pm$ pair cascade oscillations may be able to further
improve the existing constraints on axion properties.

Driven by the need to explain pulsar radio emission, semi-analytic models
motivated by first-principles simulations have been constructed
recently~\citep{Cruz:2020vfm,Tolman:2022unu}. These models improved over the
work by \citet{Levinson:2005fa} by taking into account kinetic effects. However,
no single semi-analytic model so far can properly reproduce the limit-cycle behavior of
the pair cascade process, which includes the growth of the inductive electric field
and its subsequent screening from pair production. Numerical simulations, on the
other hand, are extremely expensive when the full QED cascade physics is taken
into account, and it is impossible to simulate the parameter regime of realistic
pulsars in 2D or 3D with current computational capabilities.

In this paper, we attempt to construct a minimal time-dependent theoretical
model that can reproduce all salient features of the pulsar $e^\pm$ pair cascade
process. This model takes into account the highly relativistic and nonlinear
plasma physics governing the pair plasma near the pulsar polar cap, and uses a
physically motivated prescription to incorporate pair production and escape. In
Section~\ref{sec:equations-plasma}, we outline this theoretical model and derive
the differential equations governing it. In Section~\ref{sec:source-term}, we
discuss our choice of the pair-production source term in the equations and compare
a few different alternatives. In Section~\ref{sec:solutions}, we present
numerical solutions to this model and discuss their parameter dependence.
Finally, in Section~\ref{sec:discussion} we discuss the observational
implications and how this model can be improved in the future.

\section{Theoretical Model}
\label{sec:equations-plasma}

\subsection{Equations and Closure}
\label{sec:units}
We start from the coupled Vlasov-Maxwell system in 1D, along the local magnetic field, with a source term for pair production. Using $x$ to denote the coordinate along the field line, the system reads:
\begin{align}
  \frac{\partial E}{\partial t} &= c(\nabla\times \mathbf{B})_{\parallel} - 4\pi j \\
  \pdv{f_\pm}{t} &= - v \pdv{f_\pm}{x} - q_{\pm}E \pdv{f_\pm}{p} + S_{\pm}, \label{eqn:vlasov}
\end{align}
where $\pm$ denotes the electron or positron species, and $S_\pm$ is the source term due to $e^{\pm}$ pair production. Assuming the magnetic field is static, and does not couple to the dynamics in the spark gap, we write $j_{B} = c(\nabla\times \mathbf{B})_{\parallel}/4\pi$. Amper\'e's law becomes:
\begin{equation}
    \label{eq:ampere}
    \frac{\partial E}{\partial t} = 4\pi(j_{B} - j).
\end{equation}

The electric current density $j$ in equation~\eqref{eq:ampere} can be computed directly from the distribution function:
\begin{equation}
    \label{eq:j}
    j = \sum_{s=\pm} q_s \int v f_s\,\dd p.
\end{equation}
Taking the time-derivative of the current and using the Vlasov equation~\eqref{eqn:vlasov}, we can write down the evolution equation for $j$:
\begin{equation}
    \label{eq:djdt}
    \begin{split}
    \frac{\partial j}{\partial t} &= \sum_{s=\pm} q_s \int v \frac{\partial f_s}{\partial t}\,\dd p \\
      &= \sum_{s=\pm} q_{s} \int v\left(-v\frac{\partial f_{s}}{\partial x} - q_{s}E\frac{\partial f}{\partial p} + S_{s}\right)\,\dd p \\
      &= \sum_{s=\pm} \int \left[ q_{s}v \left(-v\frac{\partial f_{s}}{\partial x} + S_{s}\right) + q_{s}^{2}E f_{s} \frac{dv}{dp}\right]\,\dd p,
    \end{split}
\end{equation}
where we have used integration by parts to move the $\partial_{p}$ onto $v$. Since $dv/dp = 1/m\gamma^{3}$, the last term can be written as:
\begin{equation}
    \int q_{s}^{2}E f_{s}\frac{dv}{dp}\,\dd p =
    \frac{q_{s}^{2}En_{s}}{m}\left< \frac{1}{\gamma_s^{3}}\right>,
\end{equation}
where $n_{s} = \int f_{s}\dd p $ is the number density of the particle species, and the angular bracket means taking the expectation value with respect to the distribution function.


Our goal is to construct a set of coupled time-dependent ODEs that can be solved numerically. Therefore, we specialize to one point in space where particle acceleration, pair production, and electric field screening are happening, similar to the approach adopted by~\citet{Cruz:2020vfm} and~\citet{Tolman:2022unu}. To this end, we postulate a macroscopic length scale $L$ for the variation of plasma density and approximate the spatial derivative as $\partial_{x} \sim 1/L$. We further assume that pairs flow away from the point of interest at the speed of light, $v\approx c$. The Vlasov equation then becomes:
\begin{equation}
    \label{eq:vlasov-L}
    \frac{\partial f_{\pm}}{\partial t} = -\frac{c}{L}f_{\pm} - qE \pdv{f_{\pm}}{p} + S_{\pm}.
\end{equation}
This is our attempt to approximately model the plasma escape effect. Using this approximation, the time evolution equation~\eqref{eq:djdt} for the current becomes:
\begin{equation}
    \label{eq:djdt-L}
    \frac{\partial j}{\partial t} = -\frac{c}{L}j + \sum_{s=\pm}\frac{q_{s}^{2}En_{s}}{m}\left< \frac{1}{\gamma_s^{3}}\right> - q_{s}\int v S_{s}\,\dd p.
\end{equation}

Two extra quantities appear in equation~\eqref{eq:djdt-L}: the number density of each species $n_{\pm}$, and the expectation value of $1/\gamma^{3}$ for each species. The time evolution for $n_{\pm}$ can be written down by simply integrating the Vlasov equation~\eqref{eq:vlasov-L} over the momentum space:
\begin{equation}
    \label{eq:dndt}
    \frac{\partial n_{\pm}}{\partial t} = -\frac{c}{L}n_{\pm} + \int S_{\pm}\,\dd p.
\end{equation}
However, the $\langle 1/\gamma^{3}\rangle$ term requires an additional equation for closure, which invariably involves higher moments of the distribution function. In this paper, we make the simplest hydrodynamic approximation, $\langle 1/\gamma^{3}\rangle \approx 1/\langle \gamma\rangle^{3}$, and use $\langle p_{\pm}\rangle$ for closure by computing $\langle\gamma_{\pm}\rangle = \sqrt{1 + \langle p_{\pm}\rangle^{2}/m^{2}c^{2}}$. Additional discussion about this choice is included in Appendix~\ref{app:closure}, where we outline a systematic way to improve this approximation. Similar to the current equation~\eqref{eq:djdt-L}, the time evolution for $\langle p_{\pm}\rangle$ can be obtained by substituting the Vlasov equation and equation~\eqref{eq:dndt}, and then using integration by parts:
\begin{equation}
    \label{eq:dpdt}
    \begin{split}
      \frac{\partial\langle p_{\pm}\rangle}{\partial t} &= \frac{1}{n_{\pm}}\int p\frac{\partial f_{\pm}}{\partial t}\,\dd p - \frac{\dot{n}_{\pm}}{n_{\pm}}\langle p_{\pm}\rangle \\
      &= \frac{1}{n_{\pm}}\int S_{\pm}(p_{\pm} - \langle p_{\pm}\rangle)\,\dd p + q_{\pm}E.
    \end{split}
\end{equation}

Equations~\eqref{eq:djdt-L}--\eqref{eq:dpdt} involve an integral of the pair production source term $S$ which we have not specified. In this paper, we assume all pairs are produced at a single energy, but the pair production rate may be a general function of $E$, $n_{\pm}$, and $\langle p_{\pm}\rangle$, in order to take into account the feedback from electric field screening:
\begin{equation}
    \label{eq:S}
    S_{+} = S_{-} = S(E, n_{\pm}, \langle p_{\pm}\rangle)\delta(p - p_\mathrm{pair}).
\end{equation}
We will be discussing our specific choices for the function $S$ in
Section~\ref{sec:source-term}. Since electrons and positrons are accelerated in
opposite directions in the gap, the pairs produced from $\gamma$-ray photons
emitted by them have opposite momenta. In addition, the pairs produced from
curvature radiation emitted from primary particles have much lower energies than
the accelerated electrons and positrons. Therefore we set $p_\mathrm{pair} = 0$
in this model, which significantly simplifies the integrals of the source term.

\subsection{Numerical Units}
\label{sec:units}


In order to solve the coupled equations~\eqref{eq:ampere}, \eqref{eq:djdt-L},
\eqref{eq:dndt}, and~\eqref{eq:dpdt} together numerically, we introduce a unit
system that is motivated by the physics at the pulsar polar cap. In the rest of
this paper, dimensionless variables are expressed with tildes, and their units
are expressed with the subscript $0$, e.g. $\tilde{x} \equiv x/x_0$.

We normalize the electron and positron densities $n_{\pm}$ with the Goldreich-Julian density~\citep{1969ApJ...157..869G}:
\begin{equation}
    \label{eqn:ngj}
    \begin{split}
    n_\mathrm{GJ} &= \frac{\mathbf{B}\cdot \mathbf{\Omega}_\mathrm{NS}}{2\pi ec}\\
    &\simeq 6.9\times 10^{10} \left(\frac{B}{10^{12} \mathrm{\ G}}\right)\left(\frac{1\mathrm{\ s}}{P}\right)\,\mathrm{\ cm^{-3}},
    \end{split}
\end{equation}
where $B$ is the local magnetic field strength, and $\Omega$ is the rotation
angular frequency of the neutron star. The unit of length and time are
determined by the plasma frequency associated with $n_\mathrm{GJ}$:
\begin{equation}
    \label{eq:omega-gj}
    \omega_0^{2} = \frac{4\pi n_\mathrm{GJ} e^{2}}{m_{e}}.
\end{equation}
We then set the unit of time to be $t_{0} \equiv 1/\omega_{0}$ and unit of length $x_{0} \equiv c/\omega_{0}$. Using the nominal value of $n_\mathrm{GJ}$, our length unit is close to $x_{0}\approx 2\,\mathrm{cm}$. Note that the real plasma frequency after pair production sets in is going to be significantly higher, and this $\omega_{0}$ simply sets a lower bound for the plasma frequency.

For electron momentum, we set $p_0 \equiv m_e c$, and
the dimensionless momentum is simply $\tilde{p} = \gamma\beta$. The unit of the electric field $E_0 \equiv mc/et_0$ is then determined as the strength of the electric field that increases $\tilde{p}=\gamma\beta$ by $1$ in the unit time $t_0$. The unit electric current density $j_0 \equiv en_{GJ}c$ is equivalent to $j_{0}\equiv E_{0}/4\pi t_{0}$ due to our choice of $t_{0}$. Lastly, we define the units of the electron distribution function and pair production rate as $f_0 \equiv n_\mathrm{GJ}/p_0$ and $S_0 \equiv f_{0}/t_0$, respectively. In summary, our choice of units is listed below:
\begin{align*}
    t_0 &\equiv (4\pi e^2 n_\mathrm{GJ}/m_e)^{-1/2} \approx 6.7 \times 10^{-11} \mathrm{\ s}\\
    p_0 &\equiv m_ec,\
    n_0 \equiv n_\mathrm{GJ},\
    E_0 \equiv mc/et_0 \\
    j_0 &\equiv en_\mathrm{GJ}c,\
    f_0 \equiv n_\mathrm{GJ}/p_0,\ S_0 \equiv f_0/t_0. \atag
\end{align*}

After incorporating the source term~\eqref{eq:S} and using the unit listed
above, these are the equations that we solve numerically:
\begin{align}
  \label{eq:dEdt-dimless} \frac{d\tilde{E}}{d\tilde{t}} &= \tilde{j}_B - \tilde{j}, \\
  \label{eq:djdt-dimless} \frac{d\tilde{j}}{d \tilde{t}} &= -\frac{\tilde{j}}{\tilde{L}} +  \sum_{s=\pm} \left< \frac{1}{\gamma^3_{s}} \right> \tilde{E}\tilde{n}_s, \\
  \label{eq:dndt-dimless} \frac{d\tilde{n}_{\pm}}{d\tilde{t}} &= -\frac{\tilde{n}_{\pm}}{\tilde{L}} + \tilde{S}(\tilde{E},\tilde{n}_{\pm},\langle \tilde{p}_{\pm}\rangle), \\
  \label{eq:dpdt-dimless} \frac{d\langle \tilde{p}_{\pm}\rangle}{d\tilde{t}} &= -\frac{\tilde{S}(\tilde{E},\tilde{n}_{\pm},\langle \tilde{p}_{\pm}\rangle)}{\tilde{n}_{\pm}}\langle \tilde{p}_{\pm}\rangle + \frac{q_{\pm}}{e}\tilde{E}.
\end{align}
Apart from the pair production source function $\tilde{S}$ to be discussed in Section~\ref{sec:source-term}, there are two dimensionless parameters in this model, $\tilde{L}$ and $\tilde{j}_{B}$. The length scale $\tilde{L}$ parametrizes plasma escape, and will be taken to be approximately on the same order as the polar cap size $r_\mathrm{pc}$. The magnetospheric current $\tilde{j}_{B}$ provides a driving term for the electric field and is responsible for its growth after the pair plasma has advected away. We typically take $j_{B}$ to be a few times $en_\mathrm{GJ}c$, which is consistent with the magnetospheric current near the polar cap seen in global force-free and PIC simulations~\citep[see e.g.][]{2010ApJ...715.1282B,2017ApJ...851..137G}.

Combining equations~\eqref{eq:dEdt-dimless} and~\eqref{eq:djdt-dimless}, the equation for $\tilde{E}$ is essentially a damped oscillator with a constant forcing term:
\begin{equation}
    \label{eq:E-oscillator}
    \frac{d^{2}\tilde{E}}{d\tilde{t}^{2}} + \frac{1}{\tilde{L}}\frac{d\tilde{E}}{d\tilde{t}} + \sum_{s=\pm}\tilde{n}_{s}\left<\frac{1}{\gamma^{3}_{s}}\right>\tilde{E} = \frac{\tilde{j}_{B}}{\tilde{L}}.
\end{equation}
We define the effective frequency of the oscillation:
\begin{equation}
    \label{eq:omega-bar}
    \bar{\omega}^{2} = \sum_{s=\pm} \tilde{n}_{s}\left< \frac{1}{\gamma_{s}^{3}}\right>.
\end{equation}
The growth and screening of the electric field depend on the value of this
effective frequency. Initially, when $\bar{\omega}$ is very small, the
particular solution to the equation is linear growth,
$\tilde{E} \propto \tilde{j}_{B}\tilde{t}$. As the electric field increases, the
plasma is accelerated and pair production sets in, which increases
$\bar{\omega}$. The nature of the solution changes when $\bar{\omega}$ becomes
large enough to start oscillations, which initiates the electric field screening
phase.

Equation~\eqref{eq:E-oscillator} also shows that the evolution of $\tilde{E}$ and $\tilde{j}$ somewhat separates from the other two equations, and the plasma properties only affect the effective frequency in the particular combination~\eqref{eq:omega-bar}. This provides some robustness to the model that separates the dynamics of the electric field from the detailed microphysics model we use for pair production.

\section{Modeling the Pair Production Source Term}
\label{sec:source-term}

The main goal of this paper is to construct a self-contained theoretical
  model that can reproduce the limit-cycle behavior of $e^{\pm}$ pair
  production. To achieve this goal, we need to find a suitable function $S$ in
  equation~\eqref{eq:S}. The pair cascade process is a complicated (and
potentially nonlocal) sequence of curvature radiation, synchrotron radiation,
and magnetic pair production or photon-photon annihilation. Although numerical
simulations have succeeded in describing these processes accurately, it is
impossible to include their full effects in a zero-dimensional model.
Therefore, we try to find a simple qualitative source function using a trial-and-error
  process.

Since the source function $S$ may generally depend on $\tilde{E}$, $\tilde{n}_{\pm}$, and $\langle p_{\pm}\rangle$, we experimented with a number of different types of functional forms (in all cases $g$ is a dimensionless parameter):
  \begin{enumerate}[(i)]
      \item Constant pair production rate: $\tilde{S} = g$;
      \item Proportional to plasma density: $\tilde{S} = g\sum_{s}\tilde{n}_{s}$;
      \item Proportional to average momentum: $\tilde{S} = g\sum_{s}|\langle \tilde{p}_{s}\rangle|$;
      \item Proportional to local electric field: $\tilde{S} = g|\tilde{E}|$;
      \item Proportional to electric field and plasma density: $\tilde{S} = g|\tilde{E}|\sum_{s}\tilde{n}_{s}$;
      \item Proportional to average momentum and plasma density: $\tilde{S} = g\sum_{s}\tilde{n}_{s}\left|\langle \tilde{p}_{s}\rangle\right|$.
  \end{enumerate}
Type (i), constant pair production rate, is similar to the pair production function used by \citet{Tolman:2022unu}. Since it does not allow for any feedback from the evolution of $\tilde{E}$ and $\langle \tilde{p} \rangle$, we expect the gap to be screened and never grow again. It is a good choice for studying the screening phase, but we do not expect it to lead to a limit cycle. Similarly, type (ii) only depends on the plasma density, and we expect the electric field will not grow again after the initial screening phase. The rest 4 types of pair production function take into feedback from the field evolution in different ways, and it is not immediately obvious which one leads to a limit-cycle behavior.

Note that we have not included a source function type that involves a pair-production threshold $\gamma_\mathrm{thr}$, as was done by \citet{Levinson:2005fa} and \citet{Cruz:2020vfm}. We found that since the equations~\eqref{eq:dEdt-dimless}--\eqref{eq:dpdt-dimless} are already quite stiff, a threshold condition often leads to discontinuous transitions which can easily cause the solutions to run away. A more physical justification is that, since equation~\eqref{eq:dpdt-dimless} evolves the average momentum that exponentially decreases when pair production sets in, it is not appropriate to use a threshold condition on this value directly. An additional function that tracks the energies of primary particles in the gap may be better suited for applying the threshold condition, but we will leave the experimentation of such a model to future works.

\begin{figure*}[t]
    \centering
    \includegraphics[width=1\textwidth]{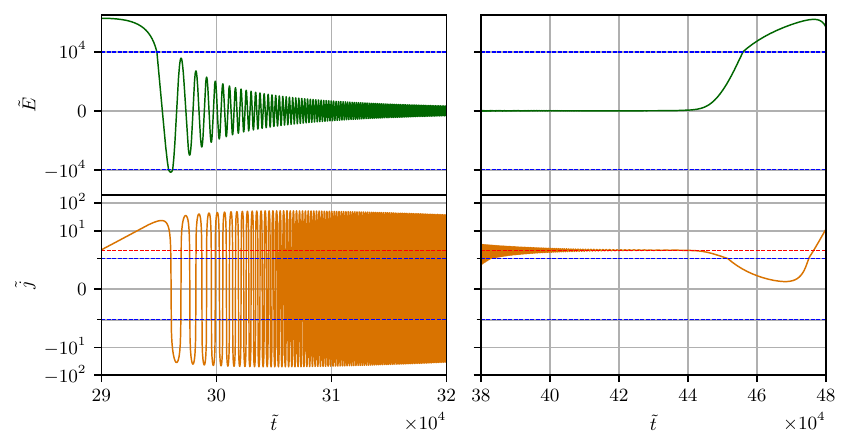}
    \caption{The zoomed-in plot of the time evolution of the electric field and
      the charged current in ``symlog'' scale, with a blue dotted line showing the threshold between linear and logarithmic scales. Left and right columns show the beginning of the screening phase and the transition to the next electric field growing phase, respectively. The escape length scale $\tilde{L}$ is set as
      $\tilde{L} = 10^4$, and pair production parameter is $g = 10^{-11}$. The value of $\tilde{j}_B$ is chosen to be $2$ and is
      shown as red dotted lines. }
    \label{fig:e-j-zoomed-in}
\end{figure*}

After extensive experimentation, we found that the only source function that can consistently reproduce the pair discharge limit-cycle behavior is type (vi), where $\tilde{S}$ is proportional to both the plasma density and the average particle momentum. A solution with this type of source function is plotted in Figure~\ref{fig:e-j-zoomed-in}, which depicts different parts of one pair discharge cycle. We will discuss this solution and its parameter dependence in Section~\ref{sec:solutions}. More discussion about how other models fail is included in Appendix~\ref{app:source term modeling}.

Coincidentally, type (vi) of the source function is applicable to pairs produced by curvature photons, which may be the primary channel for pair production at pulsar polar caps. The energy loss rate of an electron of Lorentz factor $\gamma$ undergoing curvature radiation is~\citep[see e.g.][]{jackson1999classical}:
\begin{equation}
    P_\mathrm{CR} = \frac{2}{3} \frac{e^2c}{\rho_{c}^{2}} \gamma^4,
\end{equation}
where $\rho_{c}$ is the field line curvature radius. The characteristic photon energy from curvature radiation is:
\begin{equation}
    \epsilon_\mathrm{CR} = \frac{3}{2} \frac{\lambdabar}{\rho_c} \gamma^3 m_{e}c^{2},
\end{equation}
where $\lambdabar$ is the reduced electron Compton wavelength. Therefore, the number of curvature photons $n_\mathrm{CR} \sim P_\mathrm{CR}/\epsilon_\mathrm{CR}$ emitted by an ultra-relativistic primary electron per unit time scales as $\gamma$, and each photon converts to an $e^{\pm}$ pair. Additional synchrotron cascade may increase the total number of pairs produced by a single primary electron~\citep{1983ApJ...273..761D}, but it is a reasonable first approximation to set the pair production rate to be proportional to $n_{\pm}\gamma_{\pm}$, which is our source function type (vi).

This correspondence to curvature radiation is also our way to choose the numerical parameter $g$ in the expression of the source function. The constant $g$ has the physical meaning of number of pairs produced per time $t_{0}$ per primary particle, divided by its Lorentz factor:
\begin{equation}
    \label{eq:g}
    \begin{split}
    g &= \frac{P_\mathrm{CR}}{\epsilon_\mathrm{CR}}\frac{t_{0}}{\gamma} = \frac{4}{9}\frac{e^{2}}{\lambdabar \rho_{c}m_{e}c}t_{0} \\
      &\approx 6.5\times 10^{-11}\,\left(\frac{\rho_{c}}{10^{8}\,\mathrm{cm}}\right)^{-1}B_{12}^{-1/2}P_{1}^{1/2},
    \end{split}
\end{equation}
where $B_{12} = B/10^{12}\,\mathrm{G}$ and $P_{1} = P/1\,\mathrm{s}$. For our reference model in Section~\ref{sec:solutions}, we use $g = 10^{-11}$.

\section{Numerical Solutions and Parameter Dependence}
\label{sec:solutions}

\begin{figure*}[t]
    \centering
    \includegraphics[width=1\textwidth]{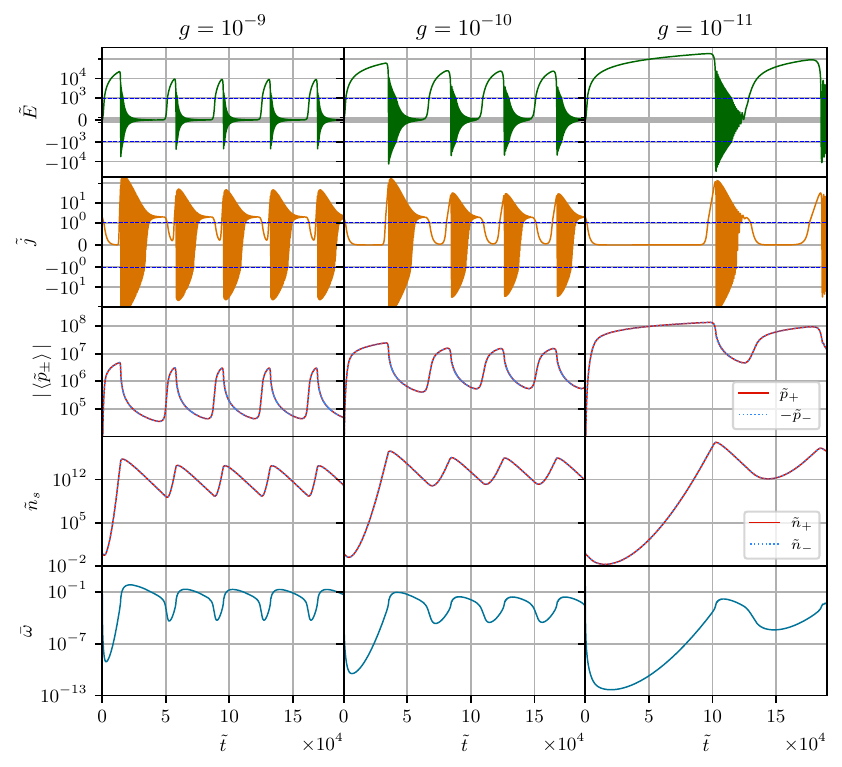}
    \caption{Full solution of equations~\eqref{eq:dEdt-dimless}--\eqref{eq:dpdt-dimless} (from the top to the bottom, $\tilde{E}, \tilde{j}, \left|\left<\tilde{p}_\pm\right>\right|$, $\tilde{n}_\pm$, and effective frequency $\bar{\omega}$) as a function of $\tilde{t}$. The first two rows are plotted in ``symlog'' scale and blue dashed lines denote the transition from linear to logarithmic scales. Each column assumes a different production rate of charged particles in pair cascades; $g = 10^{-9}, g = 10^{-10}$, and $g = 10^{-11}$ are chosen for the left, central, and right column, respectively. Initial conditions and model parameters are chosen as $\tilde{E}_\mathrm{init} = \tilde{j}_\mathrm{init} = \left<\tilde{p}_{\pm}\right>_{\mathrm{init}} = 0$, $\tilde{n}_{\pm,\mathrm{init}} = 1$, $\tilde{j}_B = 2$, and $\tilde{L} = 2\times 10^3$. The production rate of charged particles is set to be $\tilde{S} = g\sum_{\pm}\tilde{n}_{\pm}\left|\left<\tilde{p}_{\pm}\right>\right|$.}
    \label{fig: numerical results}
\end{figure*}

We solve the set of 6 coupled ODEs~\eqref{eq:dEdt-dimless}--\eqref{eq:dpdt-dimless} using the Python library function \verb+scipy.integrate.solve_ivp+. Anticipating a stiff set of equations, we use the library-provided ``Radau'' method, which is an implicit Runge-Kutta method of the Radau IIA family of order 5~\citep{hairer_solving_1993}. Since the method is adaptive, the results are interpolated through a cubic polynomial to a dense output array of at least 500,000 points, which allows for FFT calculations to analyze the power spectrum of electric field oscillations.

There are only three numerical parameters in this model. $\tilde{j}_{B}$ is
the magnetospheric current in units of $en_\mathrm{GJ}c$, which is usually order
unity at pulsar polar caps. Note that although we normalize plasma density with
$n_\mathrm{GJ}$, the system mainly operates in the regime of $\rho\sim 0$ and
the current is therefore always spacelike. $\tilde{L}$ is the dimensionless parameter
that roughly characterizes the size of the pair-producing region, and controls
how quickly plasma escapes from the region. We take it to be comparable or
smaller than the pulsar polar cap radius. The constant $g$ in the source
function $S$ parametrizes the pair production rate, and we use curvature
radiation for typical pulsar parameters to choose its value in our reference
model, as discussed in Section~\ref{sec:source-term}.

Figure~\ref{fig:e-j-zoomed-in} shows the evolution of electric field $\tilde{E}$
and current density $\tilde{j}$ over time for a particular set of parameters.
Initially, the system is filled with $n_{+} = n_{-} = n_\mathrm{GJ}$ with
  both species at rest. The electric field $E$ and electric current $j$ are
  initially also zero. Due to the magnetospheric current $j_{B}$, the electric
  field starts to grow linearly with time, accelerating both species of charges.
  Electron-positron pairs are continuously produced during this time,
  exponentially increasing the number densities $n_{\pm}$. Electric field
  screening happens when the effective frequency
  $\bar{\omega} = \sqrt{\sum \tilde{n}_{s}\langle 1/\gamma_{s}^3\rangle}$ becomes large
  enough, $\bar{\omega}\gtrsim 1/\tilde{L}$. This is when the solution
  transitions to an oscillatory nature, at which point the electric field
  screening is well-described by a weakly damped oscillator with slowly changing
  frequency. The damping effect drives the electric field to 0 and the electric current
  to $\tilde{j}_{B}$. The electric field grows again when the plasma exits the region and the effective frequency drops far below $1/\tilde{L}$.

Figure~\ref{fig: numerical results} shows the full solution for three models with different pair production parameters $g$. The electric field growth and screening limit-cycle are observed in all 3 cases. However, the three models differ in their maximum electric field strength $E_\mathrm{max}$, the quasi-period of the limit cycle, and the maximum effective frequency $\bar{\omega}_\mathrm{max}$. We see that a lower pair production efficiency leads to a larger $E_\mathrm{max}$ in the gap, and the electrons and positrons are accelerated to much higher maximum energies. The maximum effective frequency for electric field oscillations $\bar{\omega}_\mathrm{max}$ is also lower when pair production is less efficient, due to the much higher plasma Lorentz factors.

\begin{figure}
\includegraphics[width =0.5\textwidth]{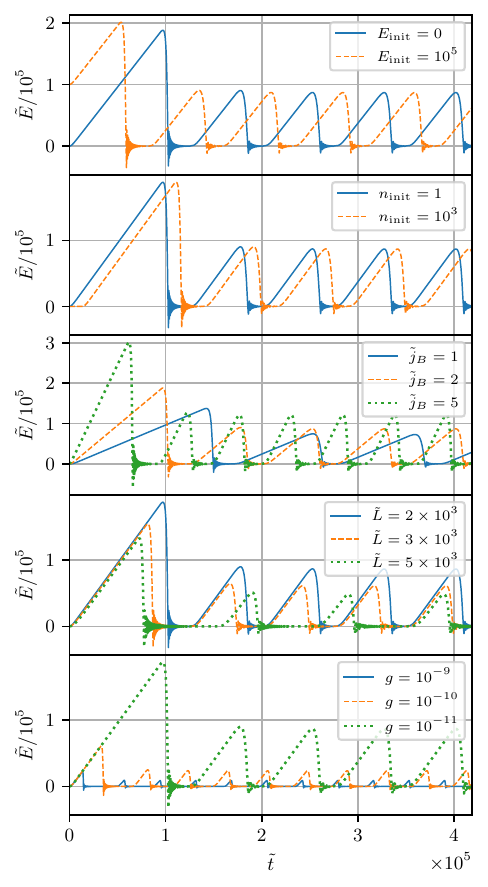}
\caption{The plots of electric fields with different choices of one of initial conditions or parameters. From the top panel, each plot compares the behavior of electric fields with several choices of $\tilde{E}_\mathrm{init}$, $\tilde{n}_\mathrm{init}$, $\tilde{j}_B$, $\tilde{L}$, and $g$, respectively. The reference model used for comparison is $\tilde{j}_{B} = 2$, $\tilde{L} = 2\times 10^{3}$, and $g = 10^{-11}$.}
\label{fig: Efield behavior}
\end{figure}

One interesting feature of this model is that, even though $\tilde{j}$
  changes sign rapidly as the electric field is screened, the mean momentum
  $\langle p_{\pm}\rangle$ just decreases but never changes sign again. We
  believe this is physical, as $\langle p_{\pm}\rangle$ can be dominated by a
  population of high energy particles, while these particles do not contribute
  as much to $\langle \beta_{\pm}\rangle$. As a result, the current density can
  become close to zero and change sign due to large amounts of pairs being
  produced. Note that $\left|\langle p_+\rangle\right|$ and $n_+$ are completely symmetric with $\left|\langle p_-\rangle\right|$ and $n_-$ in all solutions. This is simply a consequence of the symmetry $n_+\leftrightarrow n_-$ and $\langle p_+\rangle \leftrightarrow -\langle p_-\rangle$ in Equations~\eqref{eq:dndt-dimless} and \eqref{eq:dpdt-dimless}.
  
Another feature of the model is that it predicts an extremely high
  pair multiplicity of $n_{\pm}/n_\mathrm{GJ}\gtrsim 10^{16}$, far greater than
  previous more rigorous studies~\citep[e.g.][]{Timokhin:2018vdn}. This is a
  limitation of the model that stems from our choice of closure, where
  $\langle 1/\gamma_{\pm}\rangle \sim 1/\langle \gamma_{\pm}\rangle^{3}$, as well as the very simplified pair production source function discussed in Section~\ref{sec:source-term}. Our hydrodynamic closure
  may be reasonable during the electric field growing phase, where
  all particles are accelerated together, but it significantly underestimates
  this value when pair production is significant. Since electric field screening
  is controlled by the effective frequency $\bar{\omega}$, the system develops
  an unreasonably high plasma density to compensate. Experimentation with
  higher-order closures shows that a better estimate of
  $\langle 1/\gamma^{3}\rangle$ does tend to reduce the number density (see
  Appendix~\ref{app:closure}). Our source term for pair production is also quite crude, therefore we believe the model should not be used to directly study the pair multiplicities from pulsar polar cap pair cascades.

Figure~\ref{fig: Efield behavior} illustrates the dependence of the electric field solution on initial conditions and other numerical parameters. Experimenting with significantly different initial conditions, we find that the initial
condition doesn't affect the later recurrent behavior of the
solution, as all initial conditions are attracted to the same limit cycle.
Different initial electric fields and the number densities of charged particles
just result in different times of onset of the electric field screening.

\begin{figure}
\includegraphics[width=0.5\textwidth]{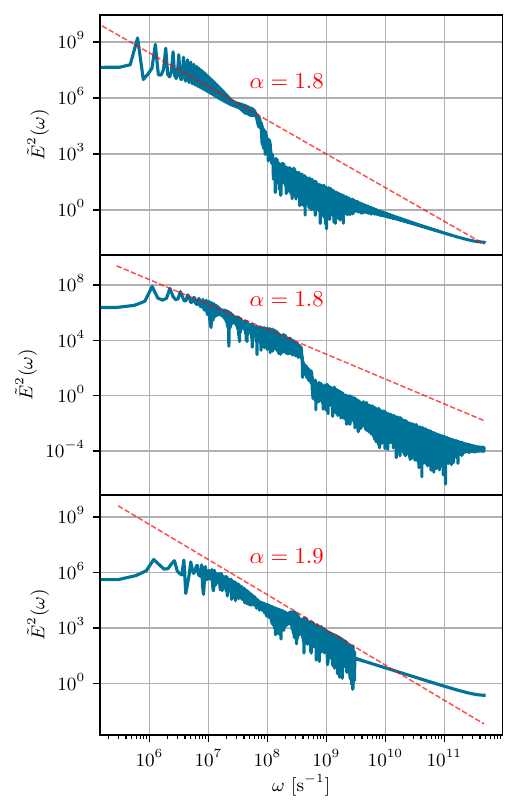}
\caption{The power spectra of the electric field $\tilde{E}^2$ for a time interval which corresponds to four pair-production limit cycles. The different plots correspond to different values of the pair production parameter: $g = 10^{-11}$ (top), $g = 10^{-10}$ (middle), and $g = 10^{-9}$ (bottom). The dotted lines are simple estimates of the spectral indices. The spectral index is only weakly dependent on the pair production rate, but the frequency cutoff depends sensitively on the parameter $g$.}
\label{fig:spectrum}
\end{figure}

On the other hand, the different choices of parameters lead to the different
time-evolution of the electric field. The growth rate of the electric field
before its screening is only governed by $\tilde{j}_B$, as that part of the
  solution is described by $\tilde{E} \propto \tilde{j}_{B}\tilde{t}$. The
systems with larger $\tilde{j}_B$ or $g$ start screening the electric fields at
the earlier time, as pair production becomes more efficient in both cases.
$\tilde{L}$ describes how slowly or quickly charged particles escape from the
pair-producing region, and thus a smaller $\tilde{L}$ leads to an
  increased duty cycle for the gap. This is because when pairs escape more
  quickly, the system spends less time until the electric fields start growing again.

This oscillating electric field in the polar cap has been suggested as a possible origin for pulsar coherent radio emission~\citep{Philippov:2020jxu}. If this is the case, then the frequency spectrum of electric field oscillations should be directly related to the resulting radio waves. Figure~\ref{fig:spectrum} shows the power
  spectrum of the electric field energy for the three models that were shown in
  Figure~\ref{fig: numerical results}. The spectrum generally follows a power law,
  but has a cutoff that scales with the pair production parameter $g$. This
  cutoff is located near the maximum $\bar{\omega}$, which is physically the
  maximum oscillation frequency of the electric field. The spectrum beyond this
  cutoff arises purely due to interpolation onto a dense output grid. The cutoff
  frequency is higher for models with more efficient pair production. For higher
  values of $g$, the frequency range contains what is typically observed in
  pulsar radio emission.

Compared to the frequency cutoff, we find that the power law index is only
  weakly dependent on the model parameters, and all lie within the $1.8$--$2.0$ range. This is largely compatible with the observed pulsar radio emission spectra of $\omega ^{-1.4 \pm 1.0}$~\citep{Bates:2013ear}. However, even within one single
  model, the spectral index seems to vary over the frequency range. More detailed
  modeling on the escape of these oscillations as coherent electromagnetic waves
  is required to make definitive predictions about the radio emission spectrum.


\section{Summary and Discussion}
\label{sec:discussion}

We have presented a minimal time-dependent theoretical model that captures the
limit-cycle behavior of the pulsar polar cap $e^{\pm}$ production process. The
model has only three parameters, and reproduces correctly the complete pair
discharge cycle, from the induction to the screening of $E_{\parallel}$. The
solution is agnostic of initial conditions, and always settles to the same
recurrent behavior under the same parameters, as characteristic of a true limit
cycle. We explored the parameter dependence of the model and computed the power
spectrum of the electric field oscillations. We found that the spectral
  index is around $1.8$--$2.0$, weakly dependent on model parameters within the
  range we have experimented with. This result is largely compatible with the
  range of spectral indices of observed pulsar radio emission.
  
In addition, the
  frequency range of the oscillations seems to be compatible with the observed
  $100\,\mathrm{MHz}$ to several GHz range. The high-frequency cutoff of the
  power spectrum depends sensitively on the efficiency of pair production, and becomes lower when pair production is less efficient. This dependence may provide an alternative explanation of the pulsar
  death line. As a pulsar spins down, its polar cap potential drops, which
  decreases the pair production efficiency. This results in an overall reduction
  of the high-frequency cutoff of its radio spectrum, rendering it undetectable
  by our radio telescopes that are sensitive between the $100\,\mathrm{MHz}$ and a few GHz.

Although the three numerical parameters in this model have straightforward
physical meanings, their values need to be fine-tuned by comparing the model
with first-principles PIC simulations such as those
by~\citet{2013MNRAS.429...20T} and~\citet{2021ApJ...919L...4C}. Such a
comparison will not only check the validity of this model, but also provide a
guide for choosing the parameters, potentially allowing for extrapolation to
realistic pulsars. Such a study will be the focus of a future work.

Our time-dependent pair production model not only describes $e^{\pm}$
discharge at the pulsar polar cap, but should also be applicable to other
systems where a spark gap is expected, e.g.\ the twisted flux tubes of a magnetar~\citep[e.g.][]{2013ApJ...777..114B} or
black hole magnetospheres~\citep[e.g.][]{2018ApJ...863L..31C}. For other systems, the pair production term may need
to be modified accordingly, but the set of
equations~\eqref{eq:dEdt-dimless}--\eqref{eq:dpdt-dimless} should generally remain
unchanged.

We have introduced a few simplifying assumptions to make the problem tractable.
One of the most significant was replacing $\langle 1/\gamma^{3}\rangle$ in
equation~\eqref{eq:djdt-dimless} by $1/\langle \gamma \rangle^{3}$, which closes
the equations but significantly underestimates the value of this term. As a result, the
number densities $n_{\pm}$ from the model are much higher than what was
generally predicted by more sophisticated theoretical models of the pair
production process~\citep[e.g.][]{Timokhin:2018vdn}. To improve the estimates of
$n_{\pm}$, more moments such as $\langle p_\pm^{2}\rangle$ can be
included in the equations as discussed in Appendix~\ref{app:closure}. However, more detailed modeling of the pair production term may also be needed in order to yield a more reliable prediction.

Another limitation of this model is that it assumes local feedback of pair
production on the electric field, which may not be realistic due to finite
photon free paths. In the case of non-local pair production, a zero-dimensional
model such as proposed in this paper is likely no longer applicable. Further
comparison with direct numerical simulations should be able to measure the
effect of non-local electric field screening and find the parameter regimes
where local pair production is a good approximation.

Despite all its limitations, we believe this model elucidates some of the most important features of the pair cascade process. In addition, this model can potentially provide a powerful way
to compute the time dependence of parallel electric field
$\mathbf{E}\cdot \mathbf{B}$ in pair producing regions in the pulsar
magnetosphere over a long period of time. Such a calculation can be used to estimate
the production rate of ALPs and their spectra. Tracing the propagation of ALPs
and their conversion back to photons can potentially give more stringent
constraints on their allowable parameter space than what is currently available.


\begin{acknowledgements}
    We thank Yajie Yuan for helpful discussions. AC is supported by NSF grants DMS-2235457 and AST-2308111. TO is supported by the U.S. Department of Energy under grant No. DE-SC 0017987.

\end{acknowledgements}


\appendix
\section{Computing the Expectation Value $\langle 1/\gamma^{3}\rangle$}
\label{app:closure}

In the main text, we have made the bold assumption that $\langle 1/\gamma^{3}\rangle \approx 1/\langle \gamma\rangle^{3}$ in order to close the equation set~\eqref{eq:dEdt-dimless}--\eqref{eq:dpdt-dimless}. In this appendix, we attempt to justify our assumption and outline a systematic way to improve this estimate. We can actually expand $1/\gamma^{3}$ around the expectation value $\gamma = \langle \gamma \rangle$:
\begin{equation}
    \label{eq:gamma3-expansion}
    \frac{1}{\gamma^{3}} = \frac{1}{\langle \gamma\rangle^{3}} - \frac{3}{\langle\gamma\rangle^{4}}(\gamma - \langle\gamma\rangle) + \frac{6}{\langle\gamma\rangle^{5}}\left(\gamma - \langle\gamma\rangle\right)^{2} + \dots
\end{equation}
This is an asymptotic expansion that may not be convergent, but it still provides us with a systematic way to approximate the expectation value of $1/\gamma^{3}$ by truncating the series at the desired order. Taking the expectation value of equation~\eqref{eq:gamma3-expansion}, the first order term goes to zero, we have:
\begin{equation}
    \label{eq:gamma3-exp-expansion}
    \left< \frac{1}{\gamma^3} \right> = \frac{1}{\left< \gamma \right>^3} + \sum^\infty_{n=2} (-1)^n\frac{(n+1)(n+2)}{2\left<\gamma\right>^{n+3}}\left< (\gamma - \left< \gamma \right>)^n \right>.
\end{equation}
Therefore, our assumption in the main text is equivalent to only keeping the zero-th order term in this expansion. This assumption is reasonable during the electric field growing phase, where all electrons and positrons are accelerated together and the distribution function has little spread, but it significantly underestimates $\langle 1/\gamma^{3}\rangle$ when pair production has begun and the electric field is screened, as shown by \citet{Cruz:2020vfm} and \citet{Tolman:2022unu}.

We can improve the estimate by including more terms in the expansion. For example, if we truncate equation~\eqref{eq:gamma3-exp-expansion} at $n=2$, the result becomes:
\begin{equation}
    \label{eq:closure-2nd}
    \left< \frac{1}{\gamma^3} \right> = \frac{1}{\left< \gamma \right>^3} + \frac{6}{\left<\gamma\right>^{5}}\left(\langle \gamma^{2}\rangle - \langle\gamma\rangle^{2}\right).
\end{equation}
At large $\gamma$, $\langle \gamma^{2}\rangle \approx \langle p^{2}\rangle$, for which we can write down an evolution equation:
\begin{equation}
    \begin{split}
      \frac{\partial \langle p^{2}\rangle}{\partial t} &= -\frac{\dot{n}}{n^{2}}\int p^{2}f\,\dd p + \frac{1}{n}\int p^{2}\frac{\partial f}{\partial t}\,\dd p \\
      &= -\frac{\dot{n}}{n}\langle p^{2}\rangle - \frac{c}{L}\langle p^{2}\rangle + 2qE\langle p\rangle + \frac{1}{n}\int p^{2}S\delta(p)\,\dd p \\
      &= -\frac{S}{n}\langle p^{2}\rangle + 2qE\langle p\rangle,
    \end{split}
\end{equation}
where we have assumed that pairs are produced at $p = 0$, as was done in the main text. If we assume a nonzero but equal and opposite $\pm p_\mathrm{pair}$, then the higher moments will contain $p_\mathrm{pair}$ as a parameter. This equation shows that the time evolution of higher moments of the distribution function depends on lower moments, and we can repeat this procedure as needed up to $n$-th order in equation~\eqref{eq:gamma3-exp-expansion}.

\begin{figure}[t]
    \centering
    \includegraphics[width=0.32\textwidth]{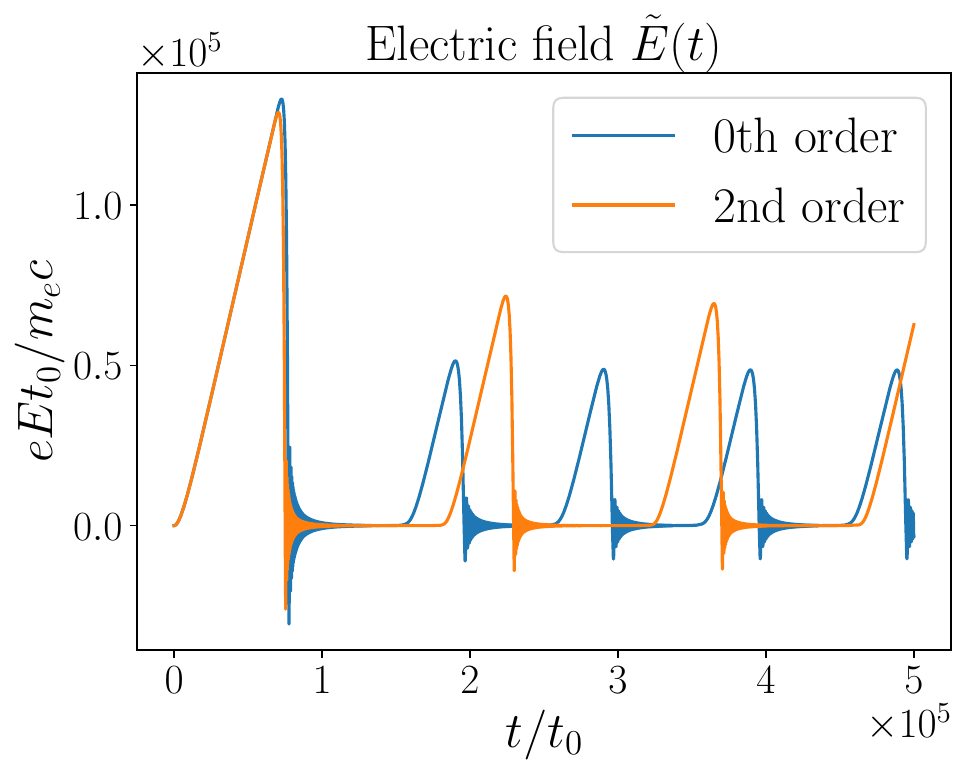}
    \includegraphics[width=0.32\textwidth]{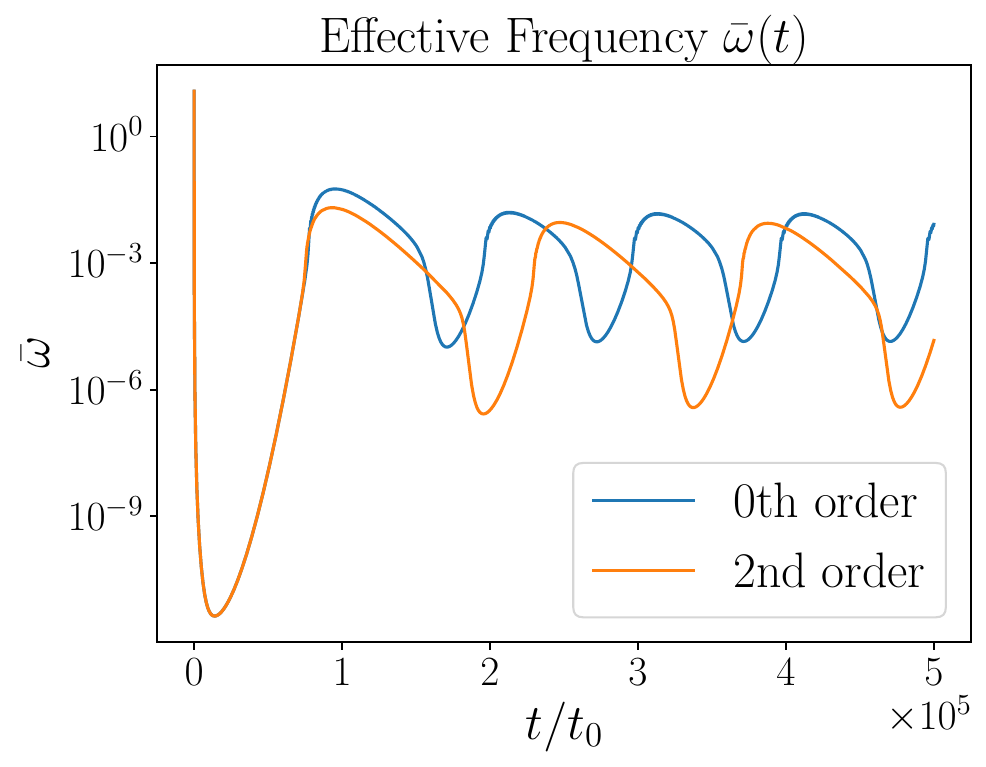}
    \includegraphics[width=0.32\textwidth]{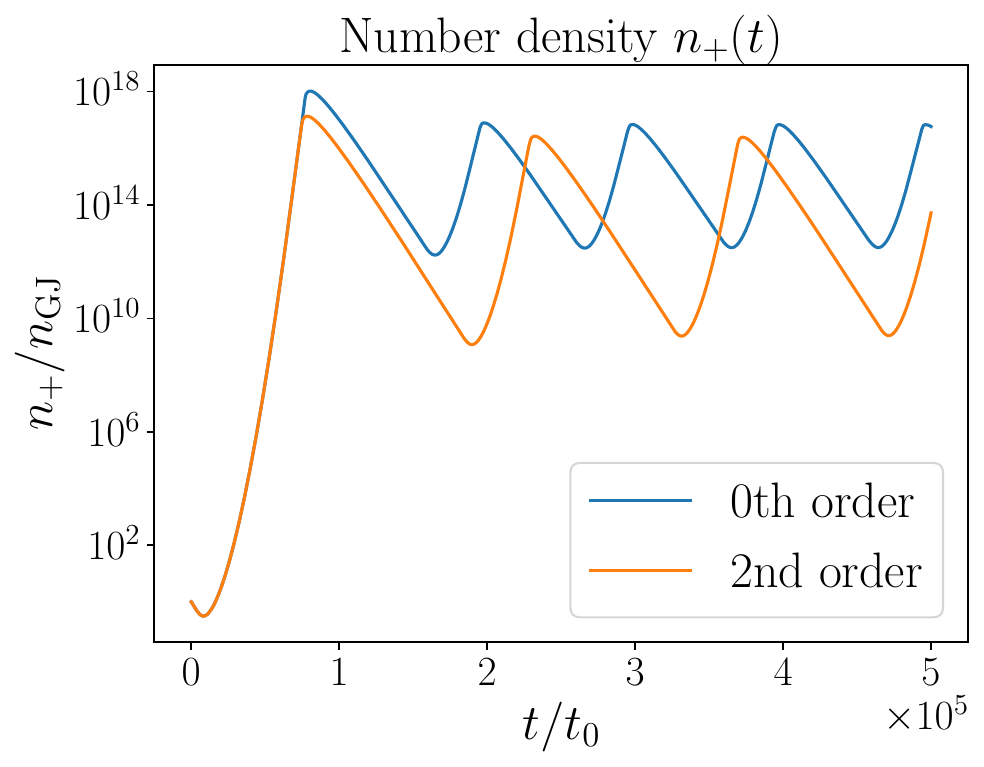}
    \caption{Comparison of the solution with different closure schemes. Blue curves show the solution with $\langle 1/\gamma^{3}\rangle = 1/\langle \gamma\rangle^{3}$, while orange curves show the solution with the second order closure defined by equation~\eqref{eq:closure-2nd}.}
    \label{fig:e-omega-compare-2nd}
\end{figure}

Figure~\ref{fig:e-omega-compare-2nd} shows a comparison between the solutions with the zeroth order closure used in the main text and the second order closure defined by equation~\eqref{eq:closure-2nd}, with all parameters identical. The overall qualitative features of the solution remain similar, but the second order closure solution has an overall lower $\bar{\omega}$. It mainly achieves this by a lower $n_{\pm}$. We expect that including more moments in the expansion will generally lead to a more realistic number density. Such a study is out of the scope of this paper and will be deferred to future works.


\section{Alternative Pair Production Models}
\label{app:source term modeling}
\begin{figure}
    \centering
    \includegraphics[width=0.49\textwidth]{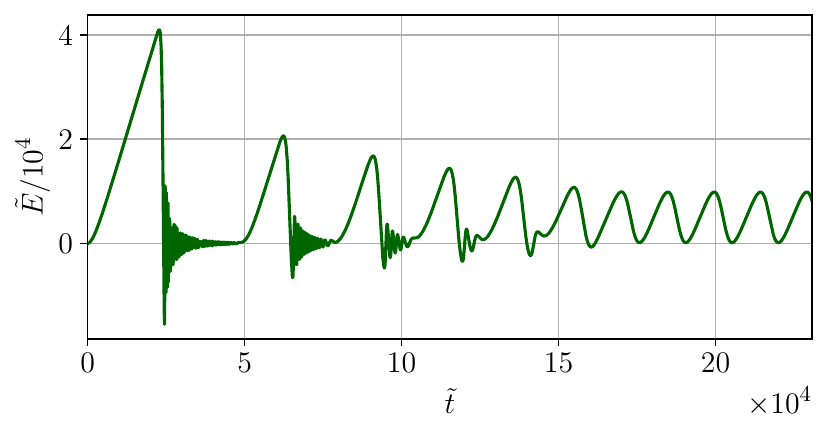}
    \includegraphics[width=0.49\textwidth]{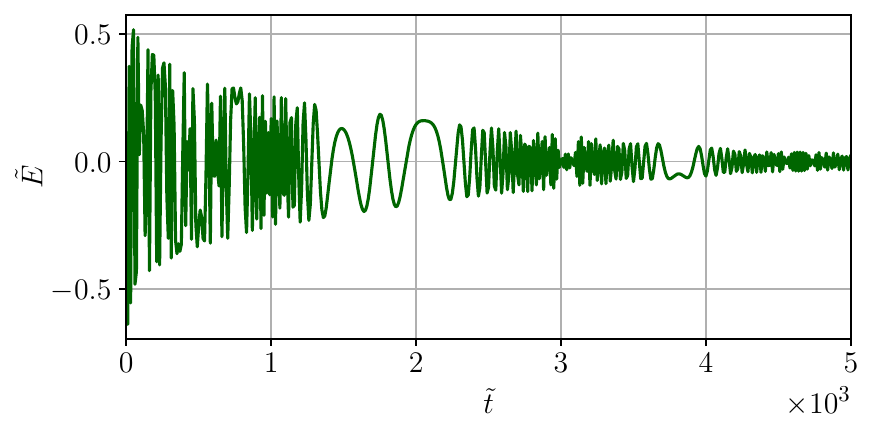}
    \caption{The plots of the electric field evolution obtained by two
      different models of the pair production source function. The left and
      right panels assume $\tilde{S} = g |\tilde{E}| \sum_s \tilde{n}_s$ and
      $\tilde{S} = g \sum_s \left|\left<\tilde{p}_s\right>\right|$, respectively. All initial
      conditions and the value of parameters are the same as the
      reference model except for the efficiency of charged particle production rate
      $g$; $g = 10^{-7}$ (left) and $g = 10^{-3}$ (right).}
    \label{fig:compare source terms}
\end{figure}

Among six models proposed in the main text, only $\tilde{S}=g \sum_s\tilde{n}_s|\left<p_s\right>|$ reproduces the limit-cycle behavior of electric fields. The other five models were not sufficient for yielding this behavior (see Figure~\ref{fig:compare source terms} for some examples) with reasonable values of initial condition and parameters. Below we briefly describe where each of the other five models was unsuccessful in reproducing the limit cycle.
\begin{itemize}
   \item $\tilde{S} = g$;

   In this case, Eq.~\eqref{eq:dndt-dimless} predicts that the deviation of the plasma density from its critical value $\tilde{n}_\pm = \tilde{L}g$ is exponentially suppressed as a function of time. The number density of charged particles in this model ends up being nearly constant.

   \item $\tilde{S} = g\sum_{s}\tilde{n}_{s}$;

   Depending on the sign of the right-hand side of Eq.~\eqref{eq:dndt-dimless}, the plasma density endlessly either grows or drops regardless of the behaviors of the electric fields, the charged current and the averaged momentum.

   \item $\tilde{S} = g\sum_{s}|\langle \tilde{p}_{s}\rangle|$;

   See Figure~\ref{fig:compare source terms}, right panel. Numerical simulations have shown that the averaged momentum tends to be relaxed to a constant value. Then, the plasma density also reaches a constant value, similar to the $\tilde{S} = \text{constant}$ case. Electric field never grows appreciably.

   \item $\tilde{S} = g|\tilde{E}|$;

   With this form of the source term $\tilde{S}$, the averaged momentum of electrons continues to decrease. That of positrons is monotonically increased or decreased depending on the value of $g$.

   \item $\tilde{S} = g|\tilde{E}|\sum_{s}\tilde{n}_{s}$;

   See Figure~\ref{fig:compare source terms}, left panel. A limit cycle tries to form but settles to an oscillation that is different in nature. In addition, Eq.~\eqref{eq:dpdt-dimless} has a particular solution in this model: $\left<\tilde{p}_s\right> = q_\pm/eg$. Thus, this model cannot describe the acceleration/deceleration of charged particles due to the electric fields.
\end{itemize}

\end{document}